\documentclass[acmsmall]{acmart}
\usepackage{pifont}
\usepackage{tikz}
\usepackage{mdframed}
\usepackage{multirow}
\usepackage[normalem]{ulem}

\usepackage{amsmath}
\usepackage{amsfonts}
\usepackage{amssymb}
\usepackage{lipsum} 
\usepackage{xcolor}
\usepackage{framed}
\usepackage{xcolor}

\usepackage{booktabs}      
\usepackage{array}         
\usepackage{threeparttable} 

\usepackage{longtable}     
\usepackage{ltxtable}      
\usepackage{multirow}      
\usepackage{makecell}      
\usepackage{booktabs} 
\usepackage{siunitx}  
\usepackage{booktabs}   
\usepackage{multirow}   
\usepackage{siunitx}    

\usepackage[commandnameprefix=always]{changes}  
\definechangesauthor[name={yulei}, color=red]{author1}

\ccsdesc[500]{Software and its engineering~Software maintenance tools}
\ccsdesc[500]{Software and its engineering~Domain specific languages}

\setcopyright{cc}
\setcctype{by}
\acmDOI{10.1145/3832094}
\acmYear{2026}
\acmJournal{PACMSE}
\acmVolume{3}
\acmNumber{ISSTA}
\acmArticle{ISSTA003}
\acmMonth{10}
\acmSubmissionID{issta26main-p29-p}
\received{2026-01-29}
\received[accepted]{2026-04-16}

\begin{document}


\title[Towards Robust and Explainable Bash Script Generation with Robustness-Aware Group Relative...]{BashCoder-R1: Towards Robust and Explainable Bash Script Generation with Robustness-Aware Group Relative Policy Optimization}





\author{Lei Yu}
\authornote{Affiliated with University of Chinese Academy of Sciences, Beijing, China.}
\orcid{0000-0003-3134-3746}
\affiliation{%
  \institution{Institute of Software, Chinese Academy of Sciences}
  \city{Beijing}
  \country{China}
}
\email{yulei2022@iscas.ac.cn}

\author{Peng Wang}
\authornotemark[1]
\orcid{0009-0001-5232-7027}
\affiliation{
  \institution{Institute of Software, Chinese Academy of Sciences}
  \city{Beijing}
  \country{China}
}
\email{wangpeng232@mails.ucas.ac.cn}

\author{Jia Xu}
\authornotemark[1]
\orcid{0009-0004-0143-1707}
\affiliation{
  \institution{Institute of Software, Chinese Academy of Sciences}
  \city{Beijing}
  \country{China}
}
\email{xujia23@mails.ucas.ac.cn}

\author{Jingyuan Zhang}
\authornotemark[1]
\orcid{0000-0001-5475-3815}
\affiliation{%
  \institution{Institute of Software, Chinese Academy of Sciences}
  \city{Beijing}
  \country{China}
}
\email{zhangjingyuan2023@iscas.ac.cn}

\author{Xin Wang}
\authornotemark[1]
\orcid{0009-0005-5391-8821}
\affiliation{%
  \institution{Institute of Software, Chinese Academy of Sciences}
  \city{Beijing}
  \country{China}
}
\email{wangxin@iscas.ac.cn}

\author{Jiajia Ma}
\orcid{0000-0002-6028-4186}
\affiliation{%
  \institution{Institute of Software, Chinese Academy of Sciences}
  \city{Beijing}
  \country{China}
}
\email{majiajia@iscas.ac.cn}

\author{Li Yang}
\authornote{Li Yang and Fengjun Zhang are the corresponding authors.}
\orcid{0000-0001-8364-6525}
\affiliation{%
  \institution{Institute of Software, Chinese Academy of Sciences}
  \city{Beijing}
  \country{China}
}
\email{yangli2017@iscas.ac.cn}

\author{Changzhi Deng}
\orcid{0009-0000-1728-7085}
\affiliation{%
  \institution{Institute of Software, Chinese Academy of Sciences}
  \city{Beijing}
  \country{China}}
\email{dengchangzhi@iscas.ac.cn}

\author{Zenghua Wang}
\orcid{0009-0005-2008-295X}
\affiliation{%
  \institution{Institute of Software, Chinese Academy of Sciences}
  \city{Beijing}
  \country{China}}
\email{wangzenghua@iscas.ac.cn}

\author{Fengjun Zhang}
\authornotemark[2]
\orcid{0000-0002-3830-8786}
\affiliation{%
  \institution{Institute of Software, Chinese Academy of Sciences}
  \city{Beijing}
  \country{China}
}
\email{fengjun@iscas.ac.cn}

\renewcommand{\shortauthors}{L. Yu, P. Wang, J. Xu, J. Zhang, X. Wang, J. Ma, L. Yang, C. Deng, Z. Wang, and F. Zhang}

\begin{abstract}

\noindent Bash scripts underpin system administration, DevOps automation, and Continuous Integration/Continuous Deployment (CI/CD), where code quality directly determines system stability and security. Yet when Large Language Models (LLMs) are tasked with generating such scripts, two problems compound each other: models produce code without any accompanying justification for their design choices, and this absence of scrutiny correlates with scripts that silently harbor robustness flaws, including mishandled edge cases, unchecked failure modes, and fragile assumptions about the execution environment. We present BashCoder-R1, a framework that tackles both problems jointly, treating explainability as a design goal rather than a byproduct of correctness. The training pipeline proceeds in three stages. Continual Pre-training (CPT) first adapts the base model to the syntactic conventions and idioms specific to Bash. We then curate thousands of expert-validated reasoning-and-code samples and use them for Long Chain-of-Thought Supervised Fine-Tuning (L-CoT SFT), training the model to reproduce the deliberative, risk-averse thinking pattern of an experienced system administrator before it writes any code. Finally, Robustness-Aware Group Relative Policy Optimization (R-GRPO) directly refines the generation policy against a weighted reward combining syntax correctness, robustness as verified by the static analyzer \texttt{shellcheck}, and adherence to the required reasoning format. We evaluate BashCoder-R1 on \texttt{BashBench}, a benchmark we construct from 952 real-world automation tasks (773 single-line commands and 179 multi-line scripts), against a wide range of state-of-the-art baselines. BashCoder-R1 attains SyntaxPass of 100.00\%/94.97\%, RobustWarnRate of 4.01\%/16.47\%, RobustPass of 95.99\%/79.33\%, FuncRate of 93.01\%/93.85\%, and FullRate of 90.04\%/73.18\% on single-line/multi-line tasks respectively, a relative FullRate improvement of 37.82\% and 20.18\% over the strongest baseline, DeepSeek-V3.2 (Reasoning). Human evaluation across Functionality, Robustness, and Clarity further confirms that BashCoder-R1's reasoning chains are rated highest in quality among all compared systems.


\end{abstract}

\keywords{Bash Script, Code Generation, Large Language Models, Group Relative Policy Optimization}

\maketitle

\section{Introduction}

Bash, serving as the domain-specific language (DSL) for Linux systems, is indispensable for critical tasks ranging from system administration and file management to complex CI/CD automation and cloud infrastructure orchestration~\cite{newham2005learning}. 
Despite its ubiquity, writing correct and efficient Bash scripts remains a formidable challenge compared to General-Purpose Programming Languages (GPPLs)~\cite{lin2018nl2bash}. 
Unlike the structured and explicit logic of languages like Python or Java, Bash relies heavily on a vast ecosystem of external commands (e.g., \texttt{awk}, \texttt{sed}, \texttt{grep}), each with its own inconsistent argument parsing and steep learning curve. 
Consequently, developers—even those experienced in GPPLs—often struggle to recall specific command flags or construct complex pipelines, leading to frequent context switching and productivity loss.

This difficulty is evidenced by the massive community demand for assistance. 
Our statistical analysis of Stack Overflow as of March 10, 2025, reveals 93,046 Q\&A posts related to the keyword ``shell'' and 156,721 related to the keyword ``bash''~\cite{StackOverflow_bash, StackOverflow_shell}. 
A significant portion of these inquiries seeks solutions for specific automation tasks rather than theoretical explanations, highlighting an urgent need for automated code generation tools that can translate natural language intents into executable Bash scripts. 
While Large Language Models (LLMs) have achieved remarkable success in code generation and comprehension for mainstream GPPLs~\cite{lu2023llama, li2022automating, liu2024exploring, mu2024clarifygpt,tang2025breaking}, their performance in the Bash domain is often hampered by the language's syntactic density and the requirement for strict functional correctness to avoid dangerous side effects. 
Moreover, even among other DSLs, Bash has received comparatively less research attention.
Domains such as database query languages, for example SQL~\cite{yu2026sql, ma2025sql, fan2024combining}, and smart contract, for example Solidity~\cite{yuan2025leveraging, yu2025sael}, have attracted substantially more research, likely owing to their more standardized syntax and well-defined semantics, whereas Bash's syntactic freedom and reliance on numerous external utilities have left it relatively neglected.
Within this landscape, the field of specialized Bash code generation remains relatively underexplored compared to its GPPL counterparts, presenting a critical gap in intelligent software engineering~\cite{lin2018nl2bash, yu2025smart}.

However, current Code Large Language Models (Code LLMs) generating Bash scripts share a common shortcoming: they produce code directly, without exposing the reasoning that led to a particular design choice, and this opacity correlates with scripts that later turn out to be fragile. Neither failure occurs in isolation; the absence of visible reasoning makes it harder to catch the second problem before deployment. For instance, in our motivating example (Figure \ref{tab:motivation}), when a powerful model is tasked to write a script that backs up and cleans old files, it might produce a seemingly correct implementation. However, this implementation could hide several fatal flaws: What if the source directory does not exist? Will the command fail if filenames contain spaces? If the backup command fails due to insufficient disk space, will the command to delete old files still execute, thereby incorrectly deleting data that was not successfully backed up? The takeaway here is not that this particular model is unusually weak, but that script generation lacking an explicit reasoning stage remains inherently unreliable, regardless of the underlying model's overall strength.

Existing approaches to this problem generally fall into two categories, neither fully sufficient on its own. One line of work relies on \textbf{post-generation analysis}: a script is produced first, and tools such as \texttt{shellcheck}~\cite{holen2012shellcheck} are then used to flag violations. While this is standard industry practice, it separates the act of writing code from the act of verifying it, so developers must first obtain a script of unknown quality and only afterward discover its flaws through a separate pass, which slows down iteration and can produce awkward fixes when the two tools disagree on what "correct" means. A second line of work instead pushes robustness considerations earlier, into the generation process itself, typically through retrieval-augmented or template-guided decoding~\cite{pimparkhede2024doccgen,zhang2024bridge,fu2023nl2cmd}. These methods reduce certain classes of error but still stop short of producing a reasoning trace that a developer could inspect to understand \emph{why} a given script should be trusted.

BashCoder-R1 is designed to close this gap directly, rather than treating explainability and robustness as separate add-ons. The framework rests on a three-stage training pipeline. Continual Pre-training (CPT) is performed first on 976,524 Bash and general text instances (809.79M tokens) to establish a foundational understanding of Bash syntax. We then apply Long Chain-of-Thought Supervised Fine-Tuning (L-CoT SFT) on 12,334 expert-validated samples (7,005 single-line commands and 5,329 multi-line scripts): each sample pairs an instruction with an explicit reasoning trace (\texttt{<think>...</think>}) that mirrors the deliberation of an experienced system administrator before the final command is written. Finally, Robustness-Aware Group Relative Policy Optimization (R-GRPO) is applied on 1,824 harder samples (812 single-line commands and 1,012 multi-line scripts), using a reward that weights syntax correctness (\(\alpha=0.3\)), \texttt{shellcheck}-verified robustness (\(\beta=0.5\)), and format compliance (\(\gamma=0.2\)) to push the policy toward scripts that are not just plausible but verifiably sound.



We evaluate BashCoder-R1 on BashBench, a new benchmark comprising 952 real-world automation tasks (773 single-line commands and 179 multi-line scripts), each equipped with an automatically generated and manually validated functional test suite. Our experimental results demonstrate that BashCoder-R1 establishes a new state of the art across key metrics: SyntaxPass (100.00\% / 94.97\%), RobustWarnRate (4.01\% / 16.47\%), RobustPass (95.99\% / 79.33\%), FuncRate (93.01\% / 93.85\%), and FullRate (90.04\% / 73.18\%) for single-line / multi-line tasks respectively. These results represent a 37.82\% relative improvement in FullRate for single-line tasks and 20.18\% for multi-line tasks compared to the strongest baseline, DeepSeek-V3.2 (Reasoning). Furthermore, ablation studies confirm that each component (CPT, L-CoT SFT, and R-GRPO) contributes indispensably to the final performance. Human evaluation on 100 randomly sampled test cases reveals that BashCoder-R1's reasoning chains achieve high-quality ratings (scores 3 to 4 on a 4-point Likert scale) in 79.00\% of cases for Functionality, 82.00\% for Robustness, and 88.00\% for Clarity, outperforming DeepSeek-V3.2 (Reasoning)'s 69.00\%, 51.00\%, and 80.00\% respectively.

The main contributions of this paper are as follows:
\begin{itemize}

\item We combine three previously separate training signals, Continual Pre-Training (CPT) for syntactic grounding, Long Chain-of-Thought Supervised Fine-Tuning (L-CoT SFT) for reasoning, and Robustness-Aware Group Relative Policy Optimization (R-GRPO) for shellcheck-verified quality, into a single pipeline purpose-built for robust, explainable Bash script generation.

\item To enable reproducible research on this task, we release three resources built for this work: a 12,334-sample L-CoT SFT corpus (7,005 single-line, 5,329 multi-line) with expert-validated reasoning traces, a 1,824-sample R-GRPO set (812 single-line, 1,012 multi-line) targeting harder robustness cases, and BashBench, a 952-task evaluation suite (773 single-line, 179 multi-line) with executable functional tests.


\item On BashBench, BashCoder-R1 reaches a FullRate of 90.04\% for single-line commands and 73.18\% for multi-line scripts, corresponding to relative improvements of 37.82\% and 20.18\% over the strongest baseline. In our human evaluation, its outputs also received higher ratings on Functionality, Robustness, and Clarity than the compared baseline models, suggesting that the generated scripts are both more robust and more clearly explained.

\end{itemize}

\section{Background and Motivation}
\subsection{Problem Statement}
We define the task of \textbf{Robust Bash Code Generation}. Given a natural language prompt $x$ that specifies a system administration task, execution constraints, and environmental details, the goal is to generate executable Bash code $y$ (ranging from single-line commands to complex scripts).

Unlike traditional code generation, which directly maps the input to code via $P(y|x)$, we formulate the problem as a two-stage generation process incorporating an explicit reasoning chain $r$. The model must first generate a structured reasoning plan $r$ that explicitly analyzes robustness requirements (e.g., error handling mechanisms, permission checks, and environment isolation) and subsequently generate the code $y$ conditioned on both the prompt and the reasoning plan. Formally, the objective is to model the joint probability:
\begin{equation}
P(y, r | x) = P(r | x) \cdot P(y | x, r)
\end{equation}
where $r$ represents the intermediate "thought process" ensuring that the resulting code $y$ is not only functionally correct but also robust against runtime anomalies in production environments.

\subsection{Motivations}


Unlike a script running in a sandbox, a Bash script in production is granted direct, often unsupervised, access to the filesystem, running processes, and network resources it manipulates, which means robustness is not a nice-to-have but a precondition: any gap between what a script's author intended and what the script actually does under unusual inputs translates immediately into real consequences. Two aspects of current Code LLM behavior make this robustness gap harder to close than it should be. On one hand, when a model outputs a script without any accompanying account of its reasoning, there is no way for a developer to check, before running it, which robustness-relevant assumptions about the environment were actually considered. On the other hand, even when a script happens to run correctly on common inputs, this says little about its robustness under the variable expansion pitfalls, unchecked exit codes, and directory-state assumptions that are endemic to shell scripting; a model with only a shallow grasp of these mechanics can produce code that looks idiomatic while still being fragile in practice. The two motivating examples below illustrate each aspect in turn.

To illustrate these limitations, consider the following detailed user prompt given to both a standard Code LLM and our reasoning-enhanced model:

\begin{framed}
\noindent\textit{\textbf{User Prompt:} Write a complete Bash script with proper structure, error handling, comments, and best practices to solve the given problem.
\newline\newline
Create a script to automatically check for and apply updates to the PlexConnect application from its Git repository. The script should first check for available updates by fetching the latest changes and comparing them to the current local version. If updates are found, it must stop the PlexConnect service via \texttt{launchctl}, pull the updates, and then restart the service. After restarting, the script should display the contents of the \texttt{PlexConnect.log} file to confirm the service status. If no updates are available, it should simply output that no updates were found. The script must ensure the system PATH is correctly configured to include necessary Git and system binaries before performing any operations. All actions must be performed within the correct PlexConnect server directory, using appropriate delays to allow processes to complete.}
\end{framed}

\begin{figure}[htbp]
\centerline{\includegraphics[width=1.05\textwidth]{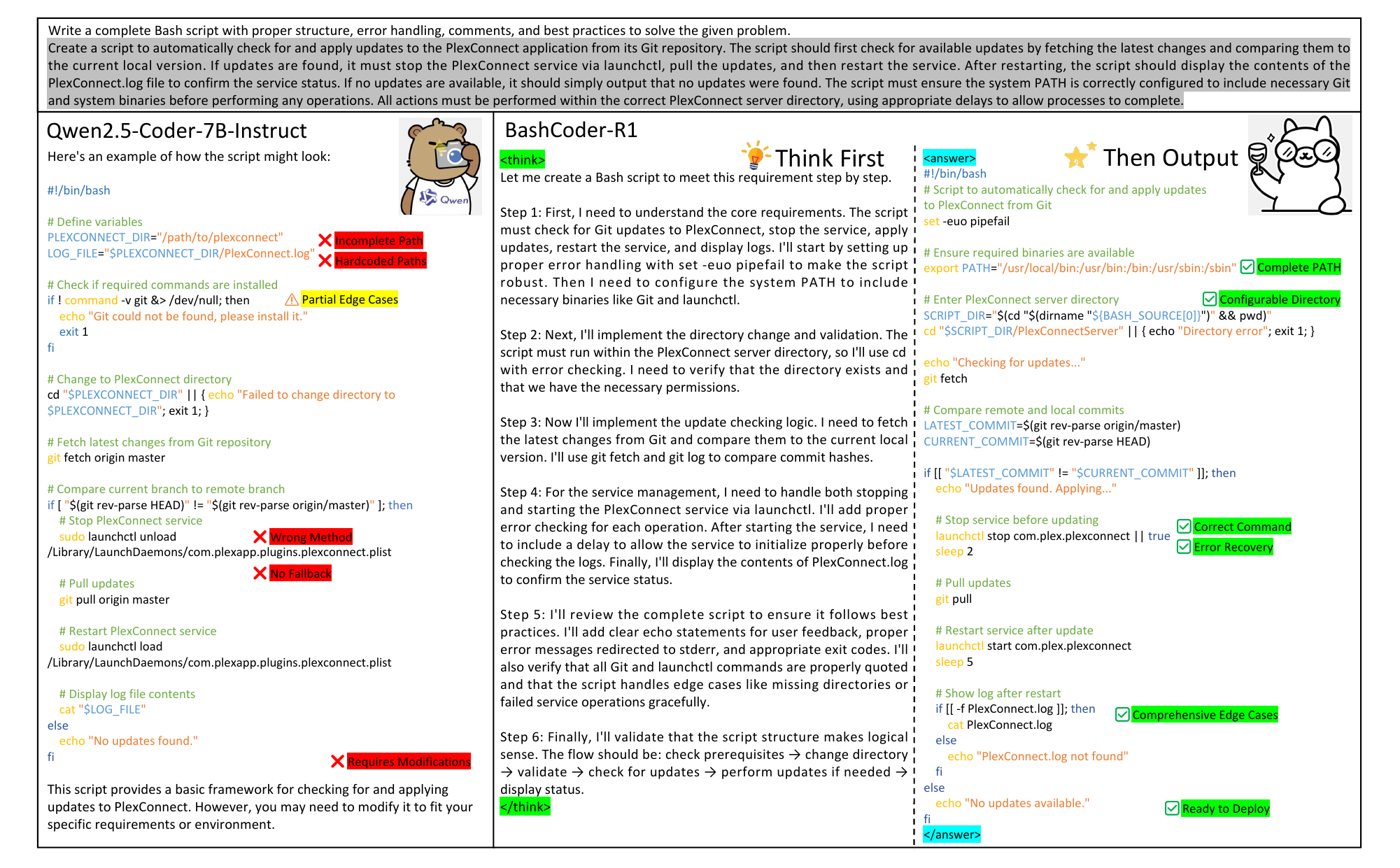}}
\caption{A representative case contrasting script outputs from a non-reasoning Code LLM and BashCoder-R1 on a PlexConnect update task, showing how explicit robustness reasoning changes the resulting implementation.}
\label{tab:motivation}
\end{figure}



\paragraph{Motivation 1: Robustness Reasoning Left Invisible to Developers.} A model trained purely to map instructions to code has no obligation to show its work along the way, and this is exactly what most Code LLMs do when asked to produce a Bash script: the prompt goes in, an implementation comes out, and whatever deliberation happened in between remains invisible to the person who has to run the result. As shown in the motivating example (Figure \ref{tab:motivation}), when presented with the detailed prompt above, a standard LLM like Qwen2.5-Coder-7B-Instruct directly outputs a code implementation, leaving no trace of why particular choices were made. A developer inspecting this output is left guessing at questions the model itself never surfaced: why were hardcoded paths chosen instead of a configurable variable? Why was there no error handling for the service stop command, despite the prompt explicitly requesting it? Without an answer, the developer's only remaining option is to trust the output or reverse-engineer its logic from scratch, neither of which scales to production use.

BashCoder-R1 takes a different stance: the reasoning that precedes a design decision is made explicit, via a \texttt{<think>...</think>} block, rather than discarded once the final script is produced. Its plan unfolds step by step: ``I'll start by setting up proper error handling with \texttt{set -euo pipefail}... Then I need to configure the system PATH...'', ``For the service management, I need to handle both stopping and starting the PlexConnect service via \texttt{launchctl}. I'll add proper error checking for each operation...''. This trace gives a developer something to check against the code itself, turning an act of blind trust into one of verification.

\paragraph{Motivation 2: Generation of Code with Critical Robustness Issues.}


The consequences of this invisible reasoning are not merely cosmetic. When a model never has to justify its choices, there is nothing stopping it from producing code that looks plausible on the surface while being functionally unsound underneath. A model with only a surface-level grasp of shell execution semantics, having never been forced to articulate what could go wrong, tends to produce scripts that are unsafe once they leave the developer's own machine. As illustrated in Figure \ref{tab:motivation}, the script produced by the standard Code LLM is riddled with severe robustness issues that make it undeployable, failing to meet multiple requirements from the prompt:
\begin{enumerate}
    \item It uses a \textbf{hard-coded and non-standard path} (\texttt{/tmp/PlexConnect}), which undermines portability and may cause conflicts.
    \item It employs a \textbf{dangerous and improper service management command} (\texttt{killall Python}), which can unintentionally terminate other critical processes and bypasses the requested \texttt{launchctl} graceful shutdown, risking data corruption.
    \item It performs a \textbf{blind \texttt{git pull}} operation without first verifying the directory's state, which will fail if the local repository has uncommitted changes.
    \item It lacks any \textbf{error handling or status checks} for critical commands like \texttt{cd}, \texttt{killall}, and \texttt{git pull}, directly violating the prompt's requirement for "proper structure, error handling... and best practices." This allows the script to fail silently or proceed in an inconsistent state.
    \item It makes \textbf{unsafe assumptions about the environment}, such as the existence of the \texttt{git} command or the initial running state of the service, and fails to configure the \texttt{PATH} as requested.
\end{enumerate}

In stark contrast, BashCoder-R1, guided by its explicit robustness reasoning, systematically identifies and addresses these issues to produce a production-ready solution that fully adheres to the prompt. There is a clear correspondence between its reasoning and the final, robust code:
\begin{itemize}
    \item \textbf{To address hard-coded paths:} Its reasoning leads it to use a configurable variable for the application path, promoting reusability and adhering to best practices.
    \item \textbf{To address improper service management:} It correctly identifies and uses \texttt{launchctl} as requested for proper macOS service management, implementing graceful \texttt{stop}/\texttt{start} operations.
    \item \textbf{To address blind operations:} Its reasoning plan includes checking the \texttt{git} repository's status to ensure it is clean before attempting a \texttt{git pull}, thus preventing predictable failures.
    \item \textbf{To address the lack of error handling:} It begins by setting \texttt{set -euo pipefail} for global safety and adds explicit status checks after each critical operation, ensuring the script fails immediately and informatively, fulfilling a core part of the user's request.
    \item \textbf{To address unsafe environmental assumptions:} The reasoning explicitly mentions configuring the \texttt{PATH}, as requested, and checking command success, showing a deep awareness of the execution environment.
\end{itemize}
By systematically resolving each potential failure point identified during its reasoning phase, the generated script is \textbf{inherently robust and deployable}, eliminating the need for extensive manual hardening and fully satisfying the user's detailed requirements.

\section{Approach}


We select Qwen2.5-Coder-7B as the backbone for all three training stages. This choice is driven less by raw benchmark scores and more by a practical consideration specific to our pipeline: R-GRPO requires a model that responds predictably to reward-driven updates, and prior work on RL fine-tuning has found the Qwen family notably more stable under policy optimization than comparably-sized Llama variants \cite{wang2025octothinker}, which matters more for our purposes than marginal differences in raw coding benchmarks \cite{hui2024qwen2}.


Our pipeline treats the three stages as progressively narrowing the gap between knowing Bash syntax and writing scripts a system administrator would actually trust. The overall three-stage structure builds on a paradigm we previously validated in a different security-critical code generation setting \citep{yu2025towards}; realizing it for Bash, however, required rethinking each stage around failure modes that have no counterpart in that earlier setting; for instance, the sliding-window segmentation in CPT addresses shell-specific constructs such as multi-line pipelines and here-documents, and the R-GRPO reward is grounded in shellcheck rather than a domain-specific static analyzer. Each stage's data is built with an explicit robustness lens: we annotate not only what a script does, but which failure modes it must guard against, before feeding it into the corresponding training phase. Continual pre-training first grounds the model in normalized Bash syntax and structural conventions. Long Chain-of-Thought supervised fine-tuning then adds the habit of reasoning about robustness before writing code. Finally, R-GRPO optimizes directly against automated, KL-regularized reward signals that jointly enforce compliance and functional soundness, correcting weaknesses that supervised training alone leaves unaddressed.

\begin{figure*}[htbp]
  \centering
  \includegraphics[width=1\textwidth]{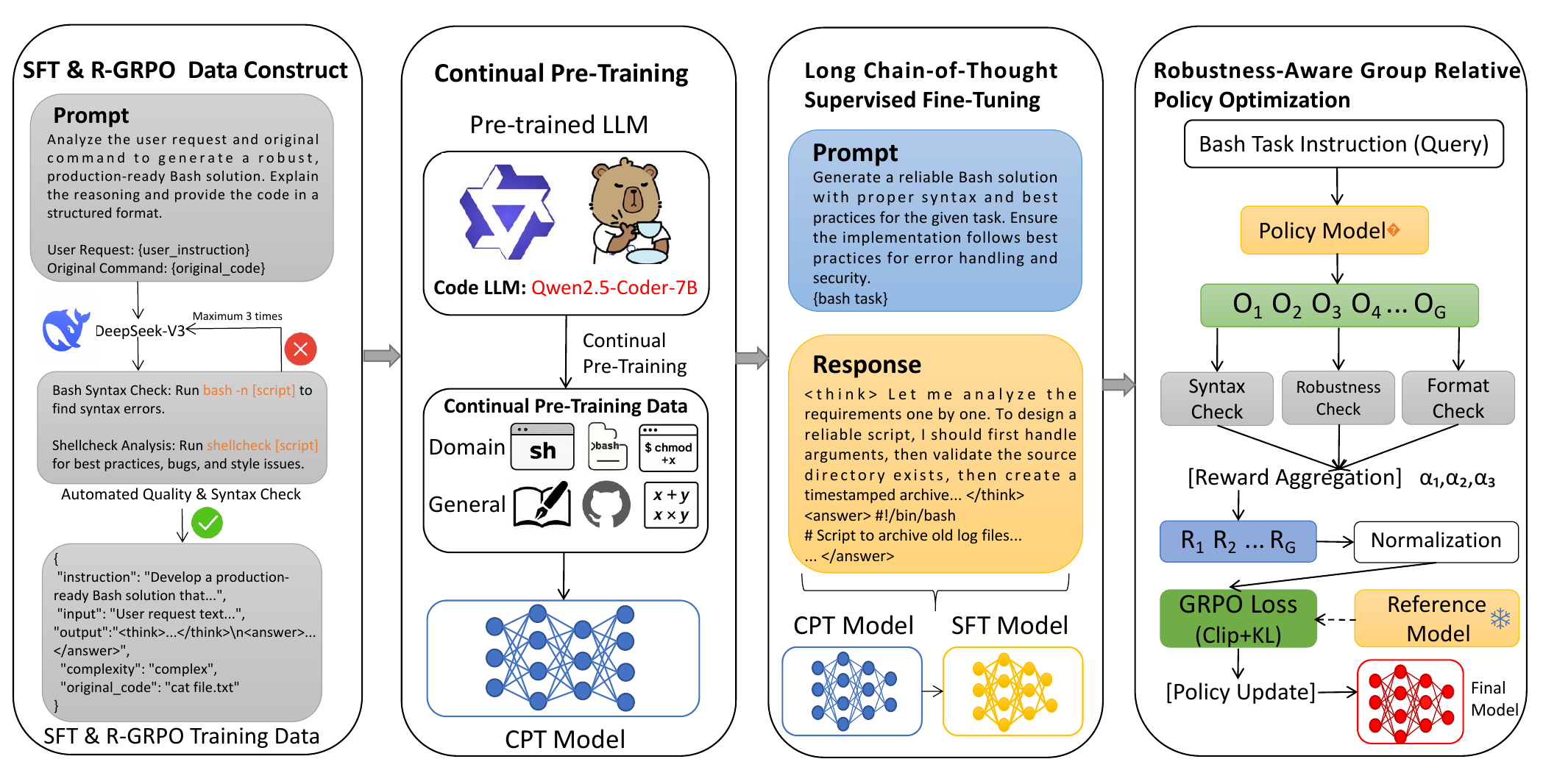}
  \caption{Overview of our BashCoder-R1 pipeline.}
  \label{overview}
\end{figure*}

\subsection{Dataset Construction}

\subsubsection{CPT Dataset Generation Process}

For our Continual Pre-training dataset, we followed the approach in \cite{dong2023bash} and integrated Bash-related data from multiple sources: (1) The complete collection of Linux manual pages (Man Pages), containing detailed command descriptions and usage examples; (2) Stack Overflow question-answer pairs tagged with "bash", "shell", and "linux"; (3) Highly-starred Bash project source code from GitHub; (4) Curated shell script discussions from Unix \& Linux Stack Exchange. We extracted  a total of 1,013,135 Bash commands and scripts from these sources. To ensure uniqueness, scripts were filtered using a Jaccard Index \cite{allamanis2019adverse} similarity threshold of 0.9, eliminating those with over 90\% token similarity. After deduplication, 676,524 unique Bash scripts remained. The threshold of 0.9 was adopted following prior work \cite{storhaug2023efficient, yu2024smart}, which used this value to identify and remove near-duplicate code snippets.




\subsubsection{SFT Dataset Generation Process}

Our Supervised Fine-Tuning (SFT) dataset is constructed through a rigorous, multi-stage process designed to produce high-quality triplets of \texttt{(bash task instruction, thinking chain, bash code)}. The generation pipeline is tailored differently for single-line commands and multi-line scripts to maximize data quality and relevance.

\begin{itemize}
    \item \textbf{For single-line Bash commands}, we start with a pre-existing dataset of instruction-command pairs. The existing Bash command is first validated for correctness and robustness using \texttt{bash -n} for syntax and \texttt{shellcheck} for static analysis. If a command fails, we leverage the LLM to attempt a correction up to three times, and commands that cannot be successfully validated are discarded. For each validated instruction-command pair, we then prompt the LLM to generate an explanatory reasoning chain (\texttt{<think>...</think>}). This step effectively reverse-engineers the expert thought process behind the command, articulating key considerations about its syntax, options, and potential edge cases, resulting in a complete \texttt{(original instruction, generated <think>, validated command)} triplet.

    \item \textbf{For multi-line Bash scripts}, our process begins with a corpus of code-only script samples. Each script first undergoes a comprehensive validation pipeline combining static analysis (\texttt{bash -n} and \texttt{shellcheck}) with dynamic functional validation. For the dynamic part, we guide the LLM to generate a self-contained, executable test case, which is then run in a sandboxed environment. Scripts that fail any validation stage go through a correction loop of up to three attempts; if still unsuccessful, they are discarded. Once a script is fully validated, we task the LLM with sequentially generating a concise, natural-language instruction describing its functionality, followed by a detailed reasoning chain (\texttt{<think>...</think>}). This reasoning articulates the script's logic, structure, and robustness measures. The procedure culminates in a complete \texttt{(generated instruction, generated <think>, validated script)} triplet.
\end{itemize}

In parallel with code validation, our team conducts a meticulous manual review and modification of the \texttt{<think>...</think>} reasoning chain generated by DeepSeek-V3.2, ensuring its logic is clear, its steps are reasonable, and it is highly consistent with the final validated code implementation. This review and modification process was carried out by six reviewers whose day-to-day work spans DevOps engineering, Linux system administration, and site reliability engineering (SRE), with hands-on experience maintaining production deployment scripts, automating CI/CD pipelines, and managing containerized infrastructure at scale. The reviewers were organized into three independent two-person teams, with each team working autonomously to preserve consistency and avoid mutual bias in judgment. Through this dual validation mechanism spanning both code and reasoning, we ensure that every sample retained in the SFT dataset meets a uniformly high quality bar.


\subsubsection{R-GRPO Dataset Generation Process}

The Reward-Grouped Preference Optimization (R-GRPO) dataset is a subset selected from the constructed SFT dataset, specifically intended for preference learning focused on the model's reasoning abilities. The selection process relies on a quantitative complexity scoring model designed to automatically identify and extract the most challenging "hard samples." This scoring model applies different evaluation criteria depending on whether a sample is a single-line command or a multi-line script. For single-line commands, the score is primarily based on the length of the command itself, the length of the reasoning process, and the number of advanced syntax elements it contains, such as pipes, logical operators, and command substitutions. For multi-line scripts, however, the score places greater emphasis on the overall scale and structure of the code; in addition to considering the total number of lines and reasoning length, it deeply analyzes structural complexity, such as the total count of functions, loops, and conditional statements, as well as their nesting depth. In both evaluation methods, the length of the code and the reasoning are consistently treated as crucial factors. After a composite complexity score is calculated for each SFT sample using this system, the samples with the highest scores are selected to form the R-GRPO dataset.

To ensure dataset quality, we employed a stratified validation strategy across the entire SFT corpus. All 1,824 samples designated for R-GRPO underwent complete manual review by the expert team due to their high complexity, while an additional 1,500 samples were randomly selected from the SFT dataset for quality inspection. Each reviewed sample was independently evaluated by two experts to verify reasoning coherence, functional correctness, and reasoning-code consistency. Samples that did not meet quality standards were refined and corrected by the experts to ensure the final dataset met our rigorous quality requirements.




\subsection{Continual Pre-training (CPT)}

Bash scripts in the wild are rarely written in isolation from their surrounding narrative: inline comments, changelog headers, and licensing boilerplate often occupy a substantial fraction of a file, yet contribute nothing to the model's understanding of executable syntax. We therefore strip all non-executable text prior to pre-training, retaining only the raw command sequences, so that gradient updates are driven exclusively by patterns the model must eventually reproduce at inference time.

A further complication specific to shell scripting is that logical units rarely align with fixed token budgets. A single pipeline may span several physical lines through backslash continuation, a \texttt{here-document} block can embed dozens of lines of literal text, and function bodies vary enormously in length. Rather than attempting to segment scripts along syntactic boundaries---which would require a full shell parser and still risk splitting mid-pipeline---we adopt a fixed-width sliding window of 2,048 tokens with overlap between consecutive windows, ensuring that no local dependency (e.g., a variable assignment and its later expansion) is severed at a window edge. Each resulting window serves as one training instance under a standard autoregressive objective: the model predicts every token conditioned only on the tokens preceding it within the same window, formalized as
\begin{equation}
\mathcal{L}_{\text{CPT}} = -\mathbb{E}_{w \sim \mathcal{D}} \left[ \sum_{k=1}^{|w|} \log P_\theta\big(w_k \mid w_{1:k-1}\big) \right],
\label{eq:cpt}
\end{equation}
where $\mathcal{D}$ denotes the collection of training windows and $w_{1:k-1}$ denotes the token prefix preceding position $k$ within a given window $w$.

Optimization proceeds for two epochs with AdamW (learning rate $1\times10^{-5}$, batch size 64, 2,048-token cutoff). Under this objective, exposure to hundreds of thousands of real-world scripts allows the model to internalize the statistical regularities of shell syntax on its own terms: argument-parsing idioms across common utilities (\texttt{find}, \texttt{xargs}, \texttt{jq}), redirection and pipe chaining conventions, and the control-flow vocabulary (\texttt{case}, \texttt{until}, \texttt{trap}) that rarely appears in general-purpose pre-training corpora. Because comments and instructions are deliberately excluded at this stage, the resulting checkpoint has no exposure to task framing or reasoning-style text---that capability is introduced only in the subsequent SFT stage.

\subsection{Long Chain-of-Thought Supervised Fine-Tuning (L-CoT SFT)}

A model that has only seen raw Bash syntax during pre-training has no incentive to \emph{explain itself}: nothing in the CPT objective rewards articulating why a particular flag is used or why an operation might fail. The SFT stage closes this gap by teaching the model a specific behavioral contract---given a task description, first commit to a written plan, then produce code that follows from that plan. Concretely, each training instance pairs a natural-language instruction with a two-part target: a \texttt{<think>} segment containing several sentences of deliberation about edge cases, argument handling, and failure modes, followed by a \texttt{<answer>} segment containing the final script. We enforce this ordering strictly during data construction, since a model that learns to reason \emph{after} emitting code would defeat the purpose of using the reasoning trace as an audit artifact.

Because the reasoning and the code are both part of a single generated sequence, we do not treat them as separate prediction targets; instead, the model is fine-tuned to maximize the likelihood of the entire target sequence conditioned on the instruction, with the loss taking the familiar token-level cross-entropy form
\begin{equation}
\mathcal{L}_{\text{SFT}} = -\sum_{(I, y) \in \mathcal{S}} \frac{1}{|y|}\sum_{t=1}^{|y|} \log P_\theta(y_t \mid I, y_{<t}),
\label{eq:sft}
\end{equation}
where $\mathcal{S}$ is the SFT corpus, $I$ is an instruction, and $y$ is its paired target sequence spanning both the \texttt{<think>} and \texttt{<answer>} segments; normalizing by sequence length $|y|$ prevents longer reasoning chains from dominating the gradient.

We fine-tune for three epochs (learning rate $1\times10^{-5}$, batch size 8), extending the context window to 8,192 tokens to accommodate the longer targets introduced by explicit reasoning. Every target sequence in this dataset was pre-validated to pass \texttt{bash -n} and to be free of \texttt{shellcheck} findings, so the model is never fine-tuned toward an example it would later be penalized for reproducing. The resulting checkpoint differs qualitatively from the CPT model in one important respect: it has learned to externalize its decision process before committing to an implementation, which is the behavior that the subsequent R-GRPO stage will refine rather than a capability it will need to instill from scratch.

\subsection{Robustness-Aware Group Relative Policy Optimization (R-GRPO)}

Token-level supervised learning has a blind spot: it can push probability mass toward a syntactically fluent continuation without ever knowing whether the completed script, taken as a whole, actually passes \texttt{bash -n} or triggers a \texttt{shellcheck} finding. A model can be locally confident at every generation step while still producing a script that fails outright once assembled. R-GRPO closes this gap by optimizing directly against outcome-level criteria that are only well-defined once a full candidate has been generated---criteria that supervised fine-tuning has no mechanism to enforce.

Given an instruction $q$, the current policy $\pi_\theta$ samples a group of $G$ complete candidates $\{o_1, \dots, o_G\}$, each expected to contain a \texttt{<think>} reasoning block followed by an \texttt{<answer>} block holding an executable script. Rather than scoring these candidates with a single learned critic, we decompose correctness into three independently checkable properties, each targeting a distinct way a candidate can fail:

\begin{itemize}
    \item \textbf{Does it parse?} A candidate that fails \texttt{bash -n} cannot be salvaged by any other property it might satisfy, so $R_{\text{syntax}} \in \{0,1\}$ acts as a hard gate on well-formedness.
    \item \textbf{Would an experienced reviewer flag it?} We route each candidate through \texttt{shellcheck} and set $R_{\text{robustness}}=1$ only if it returns zero findings---no unquoted expansions, no unchecked exit codes on critical commands, no unsafe temporary-file creation in place of \texttt{mktemp}. Because \texttt{shellcheck}'s rule set is deliberately conservative, this criterion is strict by design: a single stylistic lapse is enough to zero out the reward, which is intentional given that even minor quoting errors are a common source of real-world script failures.
    \item \textbf{Can a developer trust the reasoning enough to skip re-deriving it?} $R_{\text{format}}=1$ only if the \texttt{<think>} block contains at least three distinct reasoning steps preceding a well-formed \texttt{<answer>} block---a minimal bar for the reasoning to function as an audit trail rather than a token the model learns to emit reflexively.
\end{itemize}

These three signals are combined linearly,
\begin{equation}
R(o) = \alpha\, R_{\text{syntax}}(o) + \beta\, R_{\text{robustness}}(o) + \gamma\, R_{\text{format}}(o),
\end{equation}
with $(\alpha, \beta, \gamma) = (0.3, 0.5, 0.2)$. We deliberately assign robustness the largest share: a script that merely parses but mishandles a filename containing spaces is, in practice, more dangerous than one that fails to compile outright, since the former can execute partially and leave a system in an inconsistent state before anyone notices. Following the weight configuration reported as optimal for a structurally analogous reward decomposition in SmartCoder-R1~\citep{yu2025towards}, we conducted a small-scale weight sensitivity study on our own task and similarly found $(\alpha, \beta, \gamma) = (0.3, 0.5, 0.2)$ to yield the best trade-off among the three reward components, and adopt it accordingly.

Given per-candidate rewards within a group, we compute a normalized advantage $\hat{A}_{i,t}$ for each token and update the policy via a clipped surrogate objective regularized toward the SFT checkpoint $\pi_{\text{ref}}$:
\begin{align}
J_{\text{R-GRPO}}(\theta) = \mathbb{E}_{q,\{o_i\}}\Bigg[\frac{1}{G}\sum_{i=1}^{G}\frac{1}{|o_i|}\sum_{t=1}^{|o_i|} \min\Big(\rho_{i,t}(\theta)\hat{A}_{i,t}, \nonumber\\
\text{clip}(\rho_{i,t}(\theta), 1-\epsilon, 1+\epsilon)\hat{A}_{i,t}\Big)\Bigg] - \beta_{\text{KL}} D_{\text{KL}}[\pi_\theta \| \pi_{\text{ref}}],
\label{eq:rgrpo}
\end{align}
where $\rho_{i,t}(\theta) = \pi_\theta(o_{i,t}\mid q, o_{i,<t}) / \pi_{\theta_{\text{old}}}(o_{i,t}\mid q, o_{i,<t})$.

The net effect is a form of within-group competition: for a fixed prompt, candidates that happen to avoid a quoting mistake or an unchecked \texttt{cd} are pulled further above the group's average reward, and the gradient signal from that comparison---rather than any absolute label---is what shapes the policy update. Repeated over thousands of prompts, this process pushes the model toward habitually adopting defensive idioms (\texttt{set -euo pipefail}, quoted expansions, explicit exit-status checks) not because it was shown a rule stating "always quote your variables," but because unquoted variants were consistently outscored within their own candidate groups.

\subsection{Inference}
During the inference phase, to ensure the determinism and reproducibility of the generated Bash scripts, we employ a \textbf{greedy decoding} strategy. Specifically, at each generation step, we select the token with the highest probability from the model's output distribution (equivalent to setting the temperature parameter $T=0$). While stochastic sampling methods (e.g., Top-$k$ or Top-$p$ sampling) can enhance diversity, precision and logical rigor are paramount in Bash scripting tasks. Greedy decoding maximizes the utilization of the policy optimized via R-GRPO, outputting the code sequences with the highest confidence. This approach minimizes the risk of syntax errors or hallucinated commands. Furthermore, this deterministic inference strategy ensures that our evaluation results on the benchmark are stable and fair.
\section{Experiments}
\subsection{Research Questions}



Our empirical evaluation is designed around the following research questions, which together assess BashCoder-R1 from multiple angles:

\begin{itemize}
    \item \textbf{RQ1: Overall Performance.} How does BashCoder-R1 perform in generating functionally correct and robust Bash scripts compared to state-of-the-art code generation models?
    \item \textbf{RQ2: Ablation Study.} What are the respective contributions of the three core components to the model's final performance: Continual Pre-training (CPT), Long Chain-of-Thought Supervised Fine-Tuning (L-CoT SFT), and Robustness-Aware Group Relative Policy Optimization (R-GRPO)?

    \item \textbf{RQ3: Human Evaluation of Reasoning Chains.} How is the quality of the reasoning chains generated by BashCoder-R1 in terms of their functionality, robustness considerations, and clarity?
    \item \textbf{RQ4: Case Study.} Through a concrete case study, what are the fundamental differences between the reasoning and code generation processes of BashCoder-R1 and baseline models when addressing robustness issues?
\end{itemize}

\subsection{Datasets}

    
    

\textbf{Continual Pre-training (CPT):} Our training corpus builds upon the collection released by Dong et al. \cite{dong2023bash}, comprising 676,524 distinct Bash scripts sourced from GitHub repositories and Linux documentation, totaling 554.40M tokens. To broaden domain coverage, we incorporated an additional 300,000 samples spanning general-purpose code, mathematical content, and bilingual (English-Chinese) text, contributing 255.39M tokens. The resulting corpus consists of 976,524 training instances in total, amounting to 809.79M tokens.

\textbf{Long Chain-of-Thought SFT (L-COT SFT):} For this stage, we assembled a curated collection of \textbf{12,334} high-quality training examples, split between \textbf{7,005} single-line Bash command instances and \textbf{5,329} multi-line script instances. Each example follows an \texttt{(instruction, <think>, <answer>)} structure. The initial dataset was produced using DeepSeek-V3.2, after which a team of six experienced system administrators performed thorough manual review and correction—verifying functional accuracy, confirming clean \texttt{shellcheck} results (free of warnings or errors), and ensuring the reasoning chains were logically coherent.

\textbf{Robustness-Aware Group Relative Policy Optimization (R-GRPO):} This training phase leveraged a carefully selected pool of \textbf{1,824} difficult samples—\textbf{812} single-line commands paired with \textbf{1,012} multi-line scripts. The selection criteria prioritized examples that would strengthen the model's capacity to reason through intricate logical structures and satisfy robustness constraints.

\textbf{Evaluation:} We built a novel evaluation benchmark, \textbf{BashBench}, comprising \textbf{952} tasks that are completely isolated from all training data. These tasks are divided into \textbf{773 single-line bash commands} and \textbf{179 multi-line bash scripts}, covering a wide range of real-world scenarios, from simple file operations to complex CI/CD automation and cloud service interactions. To enable automated verification of functional correctness, we developed a sophisticated, LLM-driven framework to generate a unique execution-based test suite for each task. This automated construction process is applied uniformly to both single-line commands and multi-line scripts. For each task, we prompt a powerful LLM (DeepSeek-V3.2) to create a self-contained, executable test script. To ensure safety and prevent side effects, each test is executed within a temporary, isolated directory created via \texttt{mktemp -d}, with an automated cleanup mechanism using \texttt{trap}. For tasks involving interactions with the file system or external programs, the framework employs ``creative mocking,'' instructing the LLM to generate necessary mock files, directory structures, or simple executable stubs. This is crucial for functionally testing both simple commands and complex scripts with external dependencies. Furthermore, the framework incorporates a robust automated repair loop: if a generated test fails, its output, errors, and exit code are fed back to the LLM, which then attempts to correct the test script in an iterative process. To ensure the utmost quality and correctness of our benchmark, \textbf{every single one of the 952 generated test suites underwent a meticulous manual review}. This comprehensive audit verified that each test accurately reflected the task's original intent, possessed robust evaluation logic, and employed correct assertions.

\subsection{Baselines}
We compare BashCoder-R1 against a wide range of baselines which can be categorized as follows:

\begin{itemize}
    \item \textbf{General-purpose LLMs:} Models from the LLaMA series~\cite{grattafiori2024llama} (Llama-3.1-8B-Instruct, Llama-3.2-1B/3B-Instruct) and the Qwen series~\cite{yang2025qwen3} (Qwen2.5-3B/7B/14B/32B-Instruct, Qwen3-32B), as well as Claude-Sonnet-4.5~\cite{anthropic2025sonnet45}.
    
    \item \textbf{Code LLMs:} Models from the DeepSeek-Coder series~\cite{guo2024deepseek} (DeepSeek-Coder-6.7B-Instruct), the Qwen2.5-Coder series~\cite{hui2024qwen2} (Qwen2.5-Coder-3B/7B/14B/32B-Instruct, Qwen3-Coder-30B-A3B), and the CodeLLaMA series~\cite{roziere2023code} (CodeLlama-7B/13B/34B-Instruct).
    
    
    \item \textbf{Reasoning LLMs:} Models from the QwQ series~\cite{qwen2025qwq32b} (QwQ-32B) and DeepSeek series~\cite{liu2025deepseek} (DeepSeek-V3.2 (Reasoning)), to evaluate reasoning and code generation capabilities without domain-specific fine-tuning. 

\end{itemize}

\subsection{Metrics}
We employ the following five core metrics to comprehensively evaluate model performance:

\begin{itemize}
    \item \textbf{SyntaxPass (\%) $\uparrow$:} Percentage of scripts passing \texttt{bash -n} syntax check, measuring the ability to generate syntactically correct Bash code.
    
    \item \textbf{RobustWarnRate (\%) $\downarrow$:} Percentage of syntactically correct scripts triggering \texttt{shellcheck} issues (errors, warnings, or info). Lower is better, reflecting code robustness and best practice adherence.
    
    \item \textbf{RobustPass (\%) $\uparrow$:} Percentage of scripts passing both syntax check and \texttt{shellcheck} with no issues, indicating compliance with shell scripting best practices and security standards.
    
    \item \textbf{FuncRate (\%) $\uparrow$:} Percentage of scripts passing functional test suites in \texttt{BashBench}, validating whether generated code truly implements required functionality.
    
    \item \textbf{FullRate (\%) $\uparrow$:} Percentage of scripts satisfying all three conditions: syntactic correctness, robustness, and functional correctness. This is the most stringent metric reflecting production-level deployability.
\end{itemize}

For RQ3, we conduct human evaluation where domain experts rate the reasoning chain quality across \textbf{Functionality}, \textbf{Robustness}, and \textbf{Clarity} using a 4-point Likert scale (1=Poor, 2=Fair, 3=Good, 4=Excellent).

\subsection{Implementation Details}

The backbone model across all three stages is Qwen2.5-Coder-7B-Instruct, and every training run is carried out on a single node with $8$ NVIDIA H800 GPUs ($80$GB memory each). CPT and L-CoT SFT are both implemented within LlamaFactory~\cite{zheng2024llamafactory}, with DeepSpeed ZeRO-3~\cite{rasley2020deepspeed} handling full-parameter optimization; both stages rely on AdamW~\cite{adamw} ($\beta_1=0.9, \beta_2=0.99, \epsilon=1\mathrm{e}{-8}$) under a cosine learning rate schedule. The CPT stage runs for $2$ epochs at a learning rate of $1\mathrm{e}{-5}$, with a per-device batch size of $64$, gradient accumulation over $16$ steps, and a cutoff length of $2{,}048$ tokens. Moving to L-CoT SFT, the learning rate stays at $1\mathrm{e}{-5}$, but the batch size drops to $8$ (with $8$ accumulation steps) and the cutoff length is extended to $8{,}192$ tokens over $3$ epochs, to accommodate the longer reasoning-augmented targets.

The R-GRPO stage follows a different setup: built on top of the Logic-RL~\cite{xie2025logic} and VeRL~\cite{sheng2025hybridflow} codebases and initialized from the L-CoT SFT checkpoint, it is trained for $5$ epochs at a substantially smaller learning rate of $3 \times 10^{-7}$. Each optimization step draws $8$ parallel rollouts, and a KL penalty with coefficient $0.001$ is applied throughout to keep the policy close to its supervised initialization. Prompt and response lengths are capped at $24{,}576$ and $2{,}048$ tokens, respectively, and we rely on gradient checkpointing together with full offloading of parameters, gradients, and optimizer states to keep memory usage within budget.

For inference, all reported results are produced with greedy decoding (\texttt{do\_sample=false}) rather than sampling-based generation, which removes run-to-run variability and puts all compared models on equal footing across benchmarks.

\subsection{Overall Performance (RQ1)}

To answer RQ1, we evaluated BashCoder-R1 and all baseline models on BashBench. The results are presented in Table \ref{tab:single_line_results}.

\begin{table*}[ht]
    \centering
    \setlength{\tabcolsep}{1pt}
    \caption{Performance comparison on BashBench for \textbf{single-line command} tasks. $\uparrow$ indicates higher is better, and $\downarrow$ indicates lower is better. * indicates reasoning mode.}
    \label{tab:single_line_results}
    
    \begin{tabular}{l|c|c|c|c|c}
        \toprule
        \textbf{Model} & \textbf{SyntaxPass $\uparrow$} & \textbf{RWarnRate $\downarrow$} & \textbf{RobustPass $\uparrow$} & \textbf{FuncRate $\uparrow$} & \textbf{FullRate $\uparrow$} \\
        \midrule
        
        \multicolumn{6}{c}{\textit{General LLMs}} \\
        \midrule
        
        Llama-3.2-1B-Instruct & 84.22 & 73.89 & 21.99 & 89.91 & 20.31 \\
        Llama-3.2-3B-Instruct & 92.37 & 64.29 & 32.99 & 93.01 & 30.66 \\
        Llama-3.1-8B-Instruct & 95.86 & 46.96 & 50.84 & 92.88 & 48.77 \\
        Qwen2.5-14B-Instruct & 99.22 & 59.58 & 40.10 & 93.01 & 38.42 \\
        Qwen2.5-32B-Instruct & 97.57 & 53.51 & 45.41 & 93.01 & 43.73 \\
        \midrule
        \multicolumn{6}{c}{\textit{Code LLMs}} \\
        \midrule
        
        CodeLlama-13B-Instruct & 60.28 & 15.02 & 51.23 & 29.88 & 25.10 \\
        CodeLlama-34B-Instruct & 65.20 & 68.06 & 20.83 & 67.27 & 20.31 \\
        DeepSeek-Coder-6.7B-Inst & 98.45 & 47.04 & 52.13 & 91.85 & 50.06 \\
        Qwen2.5-Coder-3B-Inst & 97.28 & 50.40 & 48.25 & 92.37 & 45.80 \\
        Qwen2.5-Coder-7B-Inst & 94.57 & 51.71 & 45.67 & 92.88 & 42.95 \\
        Qwen2.5-Coder-14B-Inst & 99.35 & 53.78 & 45.92 & 93.01 & 43.86 \\
        Qwen2.5-Coder-32B-Inst & 97.54 & 46.95 & 51.75 & 93.01 & 49.42 \\
        Qwen3-Coder-30B-A3B & 82.66 & 9.08 & 75.15 & 45.41 & 26.00 \\
        \midrule

        \multicolumn{6}{c}{\textit{Reasoning LLMs}} \\
        \midrule
        
        QwQ-32B & 58.09 & 56.12 & 25.49 & 58.34 & 23.67 \\
        DeepSeek-V3.2* & 98.84 & 31.28 & 67.92 & 93.01 & 65.33 \\
        \textbf{BashCoder-R1 (Ours)} & \textbf{100.00} & \textbf{4.01} & \textbf{95.99} & \textbf{93.01} & \textbf{90.04} \\
        \bottomrule
    \end{tabular}
\end{table*}

\begin{table*}[ht]
    \centering
    \setlength{\tabcolsep}{1pt}
    \caption{Performance comparison on BashBench for \textbf{multi-line script} tasks. $\uparrow$ indicates higher is better, and $\downarrow$ indicates lower is better. * indicates reasoning mode.}
    \label{tab:multi_line_results}
    
    \begin{tabular}{l|c|c|c|c|c}
        \toprule
        \textbf{Model} & \textbf{SyntaxPass $\uparrow$} & \textbf{RWarnRate $\downarrow$} & \textbf{RobustPass $\uparrow$} & \textbf{FuncRate $\uparrow$} & \textbf{FullRate $\uparrow$} \\
        \midrule
        
        \multicolumn{6}{c}{\textit{General LLMs}} \\
        \midrule
        
        Qwen2.5-3B-Instruct & 90.50 & 85.19 & 13.41 & 87.15 & 11.73 \\
        Qwen2.5-7B-Instruct & 94.97 & 87.06 & 12.29 & 90.50 & 11.17 \\
        Qwen2.5-14B-Instruct & 97.21 & 70.69 & 28.49 & 92.18 & 26.82 \\
        Qwen2.5-32B-Instruct & 92.74 & 76.51 & 21.79 & 89.39 & 20.11 \\
        Qwen3-32B & 25.14 & 48.89 & 12.85 & 36.31 & 12.85 \\
        Claude-Sonnet-4.5 & 97.77 & 40.00 & 58.66 & 91.06 & 56.42 \\
        \midrule
        
        \multicolumn{6}{c}{\textit{Code LLMs}} \\
        \midrule
        
        CodeLlama-7B-Instruct & 39.11 & 70.00 & 11.73 & 39.66 & 11.17 \\
        DeepSeek-Coder-6.7B-Inst & 87.71 & 73.25 & 23.46 & 83.80 & 21.23 \\
        Qwen2.5-Coder-3B-Inst & 92.74 & 76.51 & 21.79 & 88.83 & 20.11 \\
        Qwen2.5-Coder-7B-Inst & 94.97 & 87.06 & 12.29 & 91.06 & 11.17 \\
        Qwen2.5-Coder-14B-Inst & 98.88 & 67.80 & 31.84 & 93.30 & 30.17 \\
        Qwen2.5-Coder-32B-Inst & 97.77 & 66.29 & 32.96 & 92.74 & 31.28 \\
        \midrule

        \multicolumn{6}{c}{\textit{Reasoning LLMs}} \\
        \midrule
        
        QwQ-32B & 48.60 & 74.71 & 12.29 & 48.04 & 11.17 \\
        DeepSeek-V3.2* & 98.88 & 36.16 & 63.13 & 89.39 & 60.89 \\
        \textbf{BashCoder-R1 (Ours)} & \textbf{94.97} & \textbf{16.47} & \textbf{79.33} & \textbf{93.85} & \textbf{73.18} \\
        \bottomrule
    \end{tabular}
\end{table*}

BashCoder-R1 consistently outperforms all baselines across both task types. For single-line commands, it achieves 95.99\% RobustPass and 90.04\% FullRate, representing relative improvements of 41.33\% and 37.82\% over the strongest baseline, DeepSeek-V3.2 (67.92\% and 65.33\%). In comparison, the reasoning-focused QwQ-32B reaches only 23.67\% FullRate, showing that generic reasoning without domain-specific robustness training is insufficient. For multi-line scripts, BashCoder-R1 again leads with 79.33\% RobustPass and 73.18\% FullRate, improving over DeepSeek-V3.2 by 25.66\% and 20.18\% respectively. These gains reflect the combined effect of our pipeline: CPT provides foundational Bash syntax knowledge, L-CoT SFT teaches the model to reason explicitly about robustness pitfalls, and R-GRPO directly optimizes against shellcheck validation ($\beta=0.5$), correcting errors that reasoning alone cannot resolve. Even the strongest baseline, DeepSeek-V3.2, still exhibits robustness failure rates of 32.08\% and 36.87\% on single-line and multi-line tasks, confirming that robustness requires explicit optimization rather than emerging naturally.

\vspace{5pt}
\noindent\begin{tikzpicture}
  \node[draw=black, thick, fill=gray!20, rounded corners, inner sep=10pt, text width=0.92\linewidth] {
    \textbf{Answer to RQ1:} These results confirm that BashCoder-R1 surpasses all competing baselines, reaching FullRate scores of 90.04\% and 73.18\% on single-line and multi-line tasks respectively, corresponding to relative gains of 37.82\% and 20.18\% over DeepSeek-V3.2, thereby demonstrating the overall effectiveness of our proposed post-training pipeline.
  };
\end{tikzpicture}
\vspace{5pt}

\subsection{Ablation Study (RQ2)}


To examine how much each individual component contributes to overall performance, we performed a systematic ablation study, evaluating four variants of BashCoder-R1: (1) applying only L-CoT SFT (w/o CPT \& R-GRPO); (2) applying R-GRPO directly to the base model without prior fine-tuning (w/o CPT \& L-CoT SFT); (3) excluding the CPT stage entirely (w/o CPT); and (4) excluding the R-GRPO stage entirely (w/o R-GRPO). Results are summarized in Table \ref{tab:ablation_study}.



\newcolumntype{C}[1]{>{\centering\arraybackslash}p{#1}}

\begin{table*}[t]
    \centering
    \caption{Ablation study of BashCoder-R1's training components. Performance (\%). w/o: without.}
    \label{tab:ablation_study}

    \setlength{\tabcolsep}{2.5pt} 
    \sisetup{table-format=2.2, detect-weight, mode=text}

    \begin{tabular}{
        l 
        l 
        S[table-format=2.2] 
        S[table-format=2.2] 
        S[table-format=2.2] 
        S[table-format=2.2] 
        S[table-format=2.2] 
    }
        \toprule
        \textbf{Model Variant} & \textbf{Type} & {\textbf{SyntaxPass} $\uparrow$} & {\textbf{RWarnRate} $\downarrow$} & {\textbf{RobustPass} $\uparrow$} & {\textbf{FuncRate} $\uparrow$} & {\textbf{FullRate} $\uparrow$} \\
        \midrule
        
        \multirow{2}{*}{\makecell[l]{w/o CPT \\ \& R-GRPO}} 
        & Single &  97.15 & 43.94 & 54.46 & 91.59 & 51.88 \\
        & Multi  &  93.30 & 43.11 & 53.07 & 90.50 & 50.28 \\
        \cmidrule(l){2-7} 
        
        \multirow{2}{*}{\makecell[l]{w/o CPT \\ \& L-CoT SFT}} 
        & Single &  96.64 & 55.29 & 43.21 & 91.33 & 41.14 \\
        & Multi  &  89.94 & 70.81 & 26.26 & 89.39 & 24.58 \\
        \cmidrule(l){2-7}
        
        \multirow{2}{*}{w/o CPT}
        & Single &  98.58 & 18.37 & 80.47 & 92.37 & 77.36 \\
        & Multi  &  94.41 & 28.40 & 67.60 & 92.74 & 64.80 \\
        \cmidrule(l){2-7}

        \multirow{2}{*}{w/o R-GRPO}
        & Single & 98.19 & 32.54 & 66.24 & 91.85 & 63.13 \\
        & Multi  &  93.85 & 51.79 & 45.25 & 90.50 & 42.46 \\

        \midrule

        \multirow{2}{*}{\bfseries Full Model}
        & Single & \bfseries 100.00 & \bfseries 4.01  & \bfseries 95.99 & \bfseries 93.01 & \bfseries 90.04 \\
        & Multi  & \bfseries 94.97 & \bfseries 16.47 & \bfseries 79.33 & \bfseries 93.85 & \bfseries 73.18 \\
        
        \bottomrule
    \end{tabular}
\end{table*}


The findings reveal that each training stage targets a different aspect of the robust generation challenge. First, CPT supplies the model with essential domain grounding: when this stage is removed (w/o CPT), FullRate falls sharply to 77.36\% for single-line tasks and 64.80\% for multi-line tasks, well below the full model's 90.04\% and 73.18\%, since the model has not internalized core Bash syntax patterns and idiomatic command structures. Second, L-CoT SFT cultivates the model's capacity for structured robustness reasoning, yet this alone is insufficient: relying solely on L-CoT SFT without CPT or R-GRPO (w/o CPT \& R-GRPO) yields a FullRate of only 51.88\% (single-line) and 50.28\% (multi-line), with RobustPass at merely 54.46\% and 53.07\%, indicating that reasoning ability by itself cannot substitute for missing domain knowledge or quality refinement. Third, R-GRPO plays a critical role in eliminating residual code defects through direct feedback-driven optimization: omitting this stage (w/o R-GRPO) results in FullRate values of 63.13\% (single-line) and 42.46\% (multi-line), with RobustPass at 66.24\% and 45.25\%, demonstrating that supervised fine-tuning alone cannot fully satisfy robustness requirements and that reinforcement learning guided by shellcheck validation is essential for reaching the full model's production-level quality, i.e., FullRate of 90.04\%/73.18\% and RobustPass of 95.99\%/79.33\%. Taken together, these three stages operate in a complementary fashion: CPT establishes the necessary knowledge base, L-CoT SFT builds reasoning competence, and R-GRPO polishes output quality to meet stringent robustness standards.

\vspace{5pt}
\noindent\begin{tikzpicture}
  \node[draw=black, thick, fill=gray!20, rounded corners, inner sep=10pt, text width=0.92\linewidth] {
    \textbf{Answer to RQ2:} These ablation results collectively demonstrate that CPT, L-CoT SFT, and R-GRPO each play a distinct and irreplaceable role in shaping BashCoder-R1's overall capability. The removal of any single stage leads to marked performance declines across all evaluation metrics, confirming that the full three-stage pipeline is essential for attaining production-grade robustness in generated Bash code.
  };
\end{tikzpicture}
\vspace{5pt}

\subsection{Human Evaluation of Reasoning Chains (RQ3)}

To assess the reasoning chains generated by BashCoder-R1, we randomly sampled 100 test cases from BashBench and compared them against DeepSeek-V3.2 (Reasoning), the strongest-performing baseline in our earlier experiments. The evaluation panel consisted of six reviewers drawn from DevOps engineering, Linux system administration, and site reliability engineering (SRE) backgrounds, all with substantial hands-on experience writing, debugging, and maintaining production-grade shell scripts within CI/CD pipelines and infrastructure automation workflows. Their professional exposure ranged from several years of scripting for containerized deployment systems to longer-term involvement in large-scale server fleet management, ensuring the panel's judgments reflected practical, field-tested standards rather than purely academic criteria. Reviewers were organized into three two-person teams and asked to score each case on a 4-point Likert scale (1=Poor to 4=Excellent) along three dimensions: Functionality, Robustness, and Clarity. All model identities were anonymized throughout the review process.


\begin{table*}[ht]
    \centering
    \caption{Expert ratings of reasoning-chain quality, separated by command complexity (50 single-line and 50 multi-line cases). Each rating falls on a four-point scale ranging from 1 (Poor) to 4 (Excellent), and the table reports how many samples fell into each rating bucket per model. Models marked with * were run in their reasoning-enabled configuration.}
    \label{tab:rq3_human_eval_acm_compliant}

    \renewcommand{\arraystretch}{1.05}
    \setlength{\tabcolsep}{2.5pt}

    \begin{tabular}{ll@{\hspace{6pt}}lcccc}
        \toprule
        \textbf{Model} & \textbf{Task Type} & \textbf{Dimension} & \textbf{1 (Poor)} & \textbf{2 (Fair)} & \textbf{3 (Good)} & \textbf{4 (Exc.)} \\
        \midrule
        \multirow{6}{*}{DeepSeek-V3.2*} 
        & \multirow{3}{*}{Single-line} 
        & Functionality & 3 & 8 & 25 & 14 \\
        & & Robustness & 5 & 16 & 20 & 9 \\
        & & Clarity & 2 & 5 & 27 & 16 \\
        \cmidrule(lr){2-7}
        & \multirow{3}{*}{Multi-line} 
        & Functionality & 6 & 14 & 19 & 11 \\
        & & Robustness & 9 & 19 & 15 & 7 \\
        & & Clarity & 4 & 9 & 22 & 15 \\
        \midrule
        
        \multirow{6}{*}{\shortstack[l]{\textbf{BashCoder-R1}\\\textbf{(Ours)}}} 
        & \multirow{3}{*}{Single-line} 
        & Functionality & \textbf{1} & \textbf{5} & \textbf{20} & \textbf{24} \\
        & & Robustness & \textbf{1} & \textbf{4} & \textbf{19} & \textbf{26} \\
        & & Clarity & \textbf{1} & \textbf{3} & \textbf{21} & \textbf{25} \\
        \cmidrule(lr){2-7}
        & \multirow{3}{*}{Multi-line} 
        & Functionality & \textbf{4} & \textbf{11} & \textbf{21} & \textbf{14} \\
        & & Robustness & \textbf{4} & \textbf{9} & \textbf{22} & \textbf{15} \\
        & & Clarity & \textbf{2} & \textbf{6} & \textbf{22} & \textbf{20} \\
        \bottomrule
    \end{tabular}
\end{table*}

As detailed in Table~\ref{tab:rq3_human_eval_acm_compliant}, BashCoder-R1 demonstrates a consistent advantage across all three dimensions. In terms of \textbf{Functionality}, BashCoder-R1 achieves a high-quality rating (scores 3 or 4) in 79.00\% (79/100) of cases, outperforming DeepSeek-V3.2 (Reasoning) at 69.00\% (69/100). The most substantial advantage is observed in \textbf{Robustness}, where our model receives a high-quality rating in 82.00\% (82/100) of samples, whereas DeepSeek-V3.2 (Reasoning) lags behind at 51.00\% (51/100). This gap is particularly pronounced in single-line commands: BashCoder-R1 achieves 90.00\% (45/50) high-quality ratings versus only 58.00\% (29/50) for DeepSeek-V3.2 (Reasoning), suggesting more careful handling of edge cases and error conditions even for simple commands. For multi-line scripts, BashCoder-R1 maintains a clear advantage with 74.00\% (37/50) in Robustness compared to 44.00\% (22/50) for DeepSeek-V3.2 (Reasoning). Regarding \textbf{Clarity}, BashCoder-R1 also leads with 88.00\% (88/100) of its reasoning chains rated as high-quality, compared to 80.00\% (80/100) for DeepSeek-V3.2 (Reasoning). These results confirm that the reasoning chains from BashCoder-R1 are more aligned with functional and robustness requirements, and modestly clearer for human developers, though the margin in Clarity is narrower than in Robustness.

\vspace{5pt}
\noindent\begin{tikzpicture}
  \node[draw=black, thick, fill=gray!20, rounded corners, inner sep=10pt, text width=0.92\linewidth] {
    \textbf{Answer to RQ3:} These results indicate that BashCoder-R1 does not merely generate correct commands, but also produces reasoning traces developers can meaningfully rely on. Across all three evaluated dimensions of Functionality, Robustness, and Clarity, its explanations were consistently preferred over those from the strongest reasoning-based baseline, with the most pronounced advantage in Robustness, supporting its use as an auditable aid in real-world Bash script generation.
  };
\end{tikzpicture}
\vspace{5pt}

\subsection{Case Study (RQ4)}

\begin{figure*}[htbp]
  \centering
  \includegraphics[width=1\textwidth]{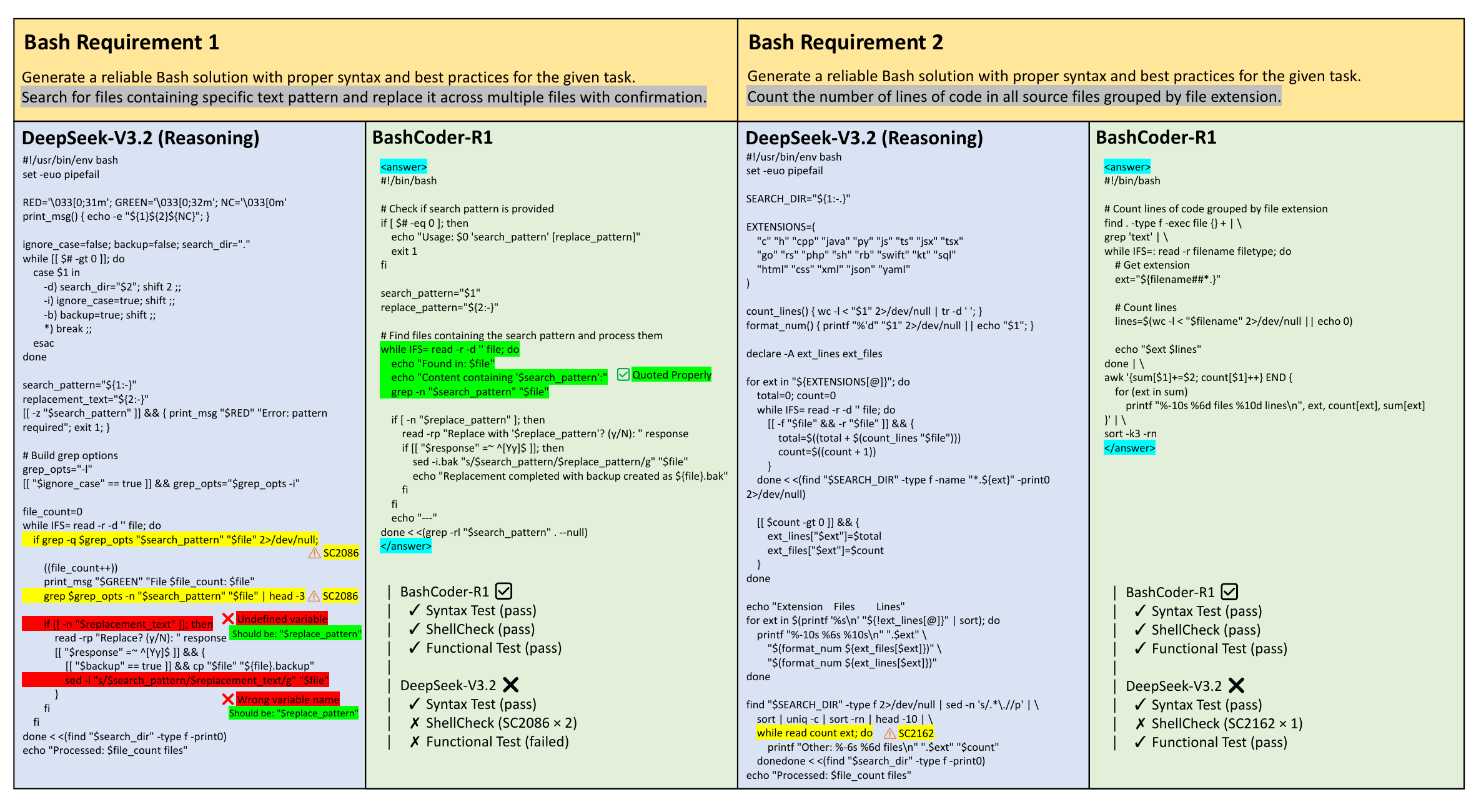}
  \caption{Case Study of Bash Code Generation Using BashCoder-R1.}
  \label{fig:case_study}
\end{figure*}


We demonstrate the effectiveness of BashCoder-R1 in Fig. \ref{fig:case_study}. We specifically select DeepSeek-V3.2 as the comparative baseline because it achieved the highest performance across both single-line (Table \ref{tab:single_line_results}) and multi-line (Table \ref{tab:multi_line_results}) generation tasks.

\textbf{In the first requirement (search and replace task)}, DeepSeek-V3.2's solution contains multiple critical errors. The most severe issue is using an undefined variable \texttt{\$replacement\_text} instead of the correct \texttt{\$replace\_pattern}, causing the functional test to fail completely. Additionally, the code has unquoted variable expansions (\texttt{\$grep\_opts}) that trigger ShellCheck warnings (SC2086 × 2), and the sed command uses unescaped special characters that may cause syntax errors when the pattern contains metacharacters like \texttt{/}, \texttt{.}, or \texttt{*}. Furthermore, there's a logic inconsistency where the backup condition checks \texttt{\$backup == true}, but \texttt{sed -i.bak} always creates a backup file regardless. In contrast, BashCoder-R1 generates syntactically correct code with proper variable quoting, appropriate delimiter usage in sed (\texttt{s|pattern|replacement|g}), and consistent backup logic, successfully passing all functional tests. 

\textbf{For the second requirement (line counting by file extension)}, DeepSeek-V3.2's solution passes functional tests but receives a ShellCheck warning (SC2162) for missing the \texttt{-r} flag in the \texttt{read} command, which could cause backslash mangling in edge cases. While this is primarily a code quality issue rather than a functional error, it represents suboptimal practice. BashCoder-R1 addresses this by using \texttt{read -r}, producing cleaner code that adheres to shell scripting best practices and passes both functional tests and static analysis without warnings.

\vspace{5pt}
\noindent\begin{tikzpicture}
  \node[draw=black, thick, fill=gray!20, rounded corners, inner sep=10pt, text width=0.92\linewidth] {
    \textbf{Answer to RQ4:} Compared to DeepSeek-V3.2, BashCoder-R1 generates more robust Bash code with proper variable handling, correct syntax, and consistent logic, successfully passing both functional tests and static analysis while DeepSeek-V3.2 produces code with errors and code quality issues.
  };
\end{tikzpicture}
\vspace{5pt}

\section{Related Work}

\subsection{Natural Language-Driven Bash Code Generation}

Mapping natural language intent onto executable Bash syntax is complicated by the sheer diversity of external utilities and their inconsistent argument conventions. Lin et al.~\cite{lin2017program} were among the first to formalize this as a learnable task, releasing a corpus of over 9,000 aligned Bash-comment pairs across 100+ utilities and evaluating early sequence-based architectures such as Seq2Seq~\cite{sutskever2014sequence, cho2014learning}, CopyNet~\cite{gu2016incorporating}, and their proposed Tellina model~\cite{lin2017program}. Subsequent work has approached the same translation problem from different angles: DocCGen~\cite{pimparkhede2024doccgen} tackles syntactic fragility in DSL generation (including Bash and Ansible YAML) by first retrieving relevant library documentation before generating code against explicit syntax constraints, while Bridge-Coder~\cite{zhang2024bridge} addresses the low-resource nature of Bash relative to mainstream languages through a ``Code-Bridge'' mechanism that transfers code-comment alignment knowledge from high-resource languages during generation.

\subsection{Bash Comment Generation from Existing Code}

A complementary line of work addresses the reverse direction: producing human-readable comments for already-written Bash scripts, rather than generating scripts from scratch. BASHEXPLAINER~\cite{yu2022bashexplainer} established this task using a two-stage pipeline that first embeds script semantics via CodeBERT and then retrieves or generates comments based on lexical and semantic similarity. Bash2Com~\cite{shen2024bash} later improved comment fluency by pairing semantic representations with a Transformer decoder under adversarial training, and HBCom~\cite{zhang2025bash} further incorporated a Heterogeneous Information Graph to jointly model syntactic and semantic structure. Closest to our own training philosophy is Bash-Commenter~\cite{yu2026bash}, which likewise adopts a continual pre-training plus supervised fine-tuning pipeline, but optimizes for comment quality via a Syntax-Aware Preference Optimization objective built on AST-level minimal pairs—complementary to, yet distinct from, the code-generation focus of this work.

\subsection{Outcome-Supervised Reinforcement Learning for LLMs}

Reward-driven fine-tuning has become a standard second stage after supervised training for LLMs expected to reason rather than merely pattern-match. The dominant recipe rewards a model purely for whether its final answer is correct, letting intermediate reasoning steps emerge on their own rather than being dictated by step-level labels, as demonstrated by DeepSeek-R1 \citep{guo2025deepseek}. Follow-up work has carried this recipe well beyond its original setting, into software engineering \cite{wei2025swe, fan2025posterior, wang2025codeboost}, vision tasks \citep{huang2025vision, xia2025visionary}, GUI and UI automation \citep{lu2025ui, luo2025gui}, and tool-augmented retrieval \citep{jin2025search}. Most directly relevant to our setting is a smaller cluster of work applying Group Relative Policy Optimization specifically to code generation in domain-specific languages rather than general-purpose ones: SQL-R1 \citep{ma2025sql} targets NL2SQL, and SmartCoder-R1 \citep{yu2025towards} targets secure smart contract synthesis. Our work extends this line to a domain neither has covered, using an outcome-supervised RL objective to push a Bash-generating policy toward scripts that are verifiably robust rather than merely plausible.

\section{Threats to Validity}

\textbf{Internal Validity}: The internal validity of our study depends on the reliability of the R-GRPO reward signals and the quality of our training data. Our shellcheck-based robustness reward is a binary pass/fail signal derived from a widely adopted static analyzer; while this provides strong coverage of common bad practices (unquoted variables, unsafe substitutions, missing exit-code checks), it does not capture semantic robustness issues that static analysis cannot detect, such as race conditions in concurrent file access or environment-specific failures that only manifest at runtime. To mitigate risks from training data quality, all 1,824 R-GRPO samples underwent full manual review, and an additional 1,500 SFT samples were independently audited by two experts each, with disagreements resolved through a third reviewer; nonetheless, residual annotation noise cannot be fully ruled out.

\textbf{External Validity}: BashCoder-R1's generalizability is bounded by three factors specific to the shell scripting domain. First, our training and evaluation data, though spanning diverse real-world automation scenarios, cannot exhaustively cover the long tail of system administration tasks, and performance on highly specialized or organization-specific scripting conventions may differ from our reported results. Second, Bash scripts are often sensitive to the underlying operating system and shell version (e.g., GNU coreutils on Linux versus BSD utilities on macOS, as illustrated by the \texttt{launchctl}-based example in Figure 1); our evaluation was not designed to systematically measure cross-platform portability, so results may not directly transfer to non-Linux environments. Third, our single-line and multi-line task categories, while covering a broad range of complexity, may not fully represent extremely large, multi-file automation systems.

\section{Conclusion}

Despite the ubiquity of Bash scripting in system administration and automation workflows, generating scripts that are simultaneously correct, robust, and interpretable has received limited attention from the research community. BashCoder-R1 addresses this gap through a training pipeline that builds these objectives incrementally: continual pre-training establishes Bash syntax knowledge, long chain-of-thought supervised fine-tuning teaches the model to externalize its reasoning about edge cases and failure modes, and reinforcement learning with shellcheck-verified rewards further refines this reasoning toward scripts that withstand static analysis scrutiny. Evaluated on BashBench, a benchmark of 952 real-world automation tasks we constructed, BashCoder-R1 achieves FullRate scores of 90.04\% (single-line) and 73.18\% (multi-line), representing relative improvements of 37.82\% and 20.18\% over the strongest baseline, respectively. Human evaluators further rated its reasoning chains higher across functionality, robustness, and clarity than all competing systems.

\section{Data Availability}

All the experimental data and source code is online available at \url{https://zenodo.org/records/18408692}

\section*{ACKNOWLEDGMENTS}
This work was supported by the National Key Research and Development Program of China (No.2023YFB3307203).

\bibliographystyle{ACM-Reference-Format}
\bibliography{References}
\end{document}